\documentclass[copyright,creativecommons]{eptcs}
\providecommand{\event}{GRAPHITE 2014} 

\usepackage{amssymb}
\usepackage{amsmath}
\usepackage{mathrsfs}
\usepackage{stmaryrd}
\usepackage{graphicx}
\usepackage{prooftree}
\usepackage[utf8]{inputenc}
\usepackage{subfigure}
\usepackage{url}

\newtheorem{theorem}{Theorem}[section]
\newtheorem{lemma}[theorem]{Lemma}
\newtheorem{proposition}[theorem]{Proposition}

\newtheorem{property}[theorem]{Property}

\newenvironment{proof}[1][Proof]{\begin{trivlist}
\item[\hskip \labelsep {\bfseries #1}]}{\end{trivlist}}
\newenvironment{definition}[1][Definition]{\begin{trivlist}
\item[\hskip \labelsep {\bfseries #1}]}{\end{trivlist}}

\newcommand{\qed}{\nobreak \ifvmode \relax \else
      \ifdim\lastskip<1.5em \hskip-\lastskip
      \hskip1.5em plus0em minus0.5em \fi \nobreak
      \vrule height0.75em width0.5em depth0.25em\fi}

\usepackage{xspace}

\newcommand\putnline[3]
{\count0 = #2 \advance \count0 by 10%
\put(#1,#2){\line(0,1){#3}}%
\put(#1,\count0){\makebox(0,0){$/$}}}

\newcommand\putaxiom[3]
{\count2 = #3 \count0 = #3 \divide \count0 by 2
\count1 = #1 \advance \count1 by \count0
\put(\count1,#2){\line(-1,0){#3}}%
\put(\count1,#2){\line(0,-1){10}}%
\advance \count1 by -\count2%
\put(\count1,#2){\line(0,-1){10}}%
}

\newcommand\putcut[3]
{\count2 = #3 \count0 = #3 \divide \count0 by 2
\count1 = #1 \advance \count1 by \count0
\put(\count1,#2){\line(-1,0){#3}}%
\put(\count1,#2){\line(0,1){10}}%
\advance \count1 by -\count2%
\put(\count1,#2){\line(0,1){10}}%
}

\newcommand\putdtriangle[3]{\count0=#2 \advance \count0 by 9%
\count1=#1 \advance \count1 by -10%
\count2=#2 \advance \count2 by 15%
\put(#1,#2){\line(-2,3){10}}%
\put(#1,#2){\line(2,3){10}}%
\put(\count1,\count2){\line(1,0){20}}%
\advance \count1 by 4%
\put(\count1,\count2){\line(0,1){10}}%
\advance \count1 by 12%
\put(\count1,\count2){\line(0,1){10}}%
\put(#1,#2){\vector(0,-1){10}}%
\put(#1,\count0){\makebox(0,0){#3}}}

\newcommand{\sema}[1] {\ensuremath{\left [ #1\right ]}\xspace} 

\newcommand{\Id}{{\sf Id}}
\newcommand{\Fail}{{\sf Fail}}

\newcommand{\CrtGraph}{{\tt CrtGraph}}
\newcommand{\CrtPos}{{\tt CrtPos}}
\newcommand{\CrtBan}{{\tt CrtBan}}

\newcommand{\AllNgb}[1]{{\tt AllNgb(}#1{\tt )}}

\newcommand{\OneNgb}[1]{{\tt OneNgb(}#1{\tt )}}
\newcommand{\NextNgb}[1]{{\tt NextNgb(}#1{\tt )}}

\newcommand{\Property}[2]{{\tt Property(}#1,#2{\tt )}}

\newcommand{\one}[1]{{\tt one(}#1{\tt )}}

\newcommand{\all}[1]{{\tt all(}#1{\tt )}}

\newcommand{\ppick}[1]{{\tt ppick(}#1{\tt )}}
\newcommand{\while}[1]{{\tt while(}#1{\tt )}}

\newcommand{\whiledoo}[2]{{\tt while(}#1{\tt )do(}#2{\tt )}}

\newcommand{\Repeat}[1]{{\tt repeat(}#1{\tt )}}

\newcommand{\orelse}[2]{(#1) {\tt orelse} (#2)}

\newcommand{\ifthenelsee}[3]{{\tt if(}#1{\tt )then(}#2{\tt )else(}#3{\tt )}}

\newcommand{\isEmpty}[1]{{\tt isEmpty(}#1{\tt )}}
\newcommand{\setPos}[1]{{\tt setPos(}#1{\tt )}}
\newcommand{\setBan}[1]{{\tt setBan(}#1{\tt )}}

\newcommand{\nott}[1]{{\tt not(}#1{\tt )}}

\newcommand{\ra}{\rightarrow}
\newcommand{\Ra}{\Rightarrow}

\newcommand{\lra}{\longrightarrow}

\newcommand{\R}{{\cal R}}

\newcommand{\Interface}{\mathit{Interface}}

\ifx\thm\undefined
\newtheorem{thm}{Theorem}
\fi

\ifx\exmp\undefined
\newtheorem{exmp}[thm]{Example}
\fi

\ifx\proposition\undefined
\newtheorem{proposition}[thm]{Proposition}
\fi

\ifx\lemma\undefined
\newtheorem{lemma}[thm]{Lemma}
\fi

\ifx\definition\undefined
\newtheorem{definition}[thm]{Definition}
\fi

\title{Strategic Port Graph Rewriting: \\
An Interactive Modelling and Analysis Framework\thanks{Partially supported by the French National Research Agency project EVIDEN (ANR 2010 JCJC 0201 01).}}

\author{Maribel Fern\'andez\institute{King's College London, Department of Informatics, Strand, London WC2R 2LS, UK}\email{maribel.fernandez@kcl.ac.uk}
\and H\'el\`ene Kirchner\institute{Inria, Domaine de Voluceau, Rocquencourt BP 105, 78153 Le Chesnay Cedex, France }\email{helene.kirchner@inria.fr}
\and Bruno Pinaud\institute{Bordeaux University, LaBRI CNRS UMR5800, 33405 Talence Cedex, France}\email{bruno.pinaud@labri.fr}
}

\def\titlerunning{Strategic port graph rewriting}
\def\authorrunning{M. Fern\'andez, H. Kirchner and B. Pinaud}

\begin{document}

\maketitle             

\begin{abstract} 
We present strategic port graph rewriting as a basis for the
implementation of visual modelling and analysis tools. The goal is to
facilitate the specification, analysis and simulation of complex
systems, using port graphs.  A system is represented by an initial
graph and a collection of graph rewriting rules, together with a
user-defined strategy to control the application of rules.  
The strategy language includes constructs to deal with 
graph traversal and management of rewriting positions in the graph.
We give a small-step operational semantics for the
language, and describe its implementation in the graph transformation
and visualisation tool PORGY.
\end{abstract}


\section{Introduction} 

\begin{figure}[ht]
\centering
\includegraphics[width=0.8\columnwidth, keepaspectratio]{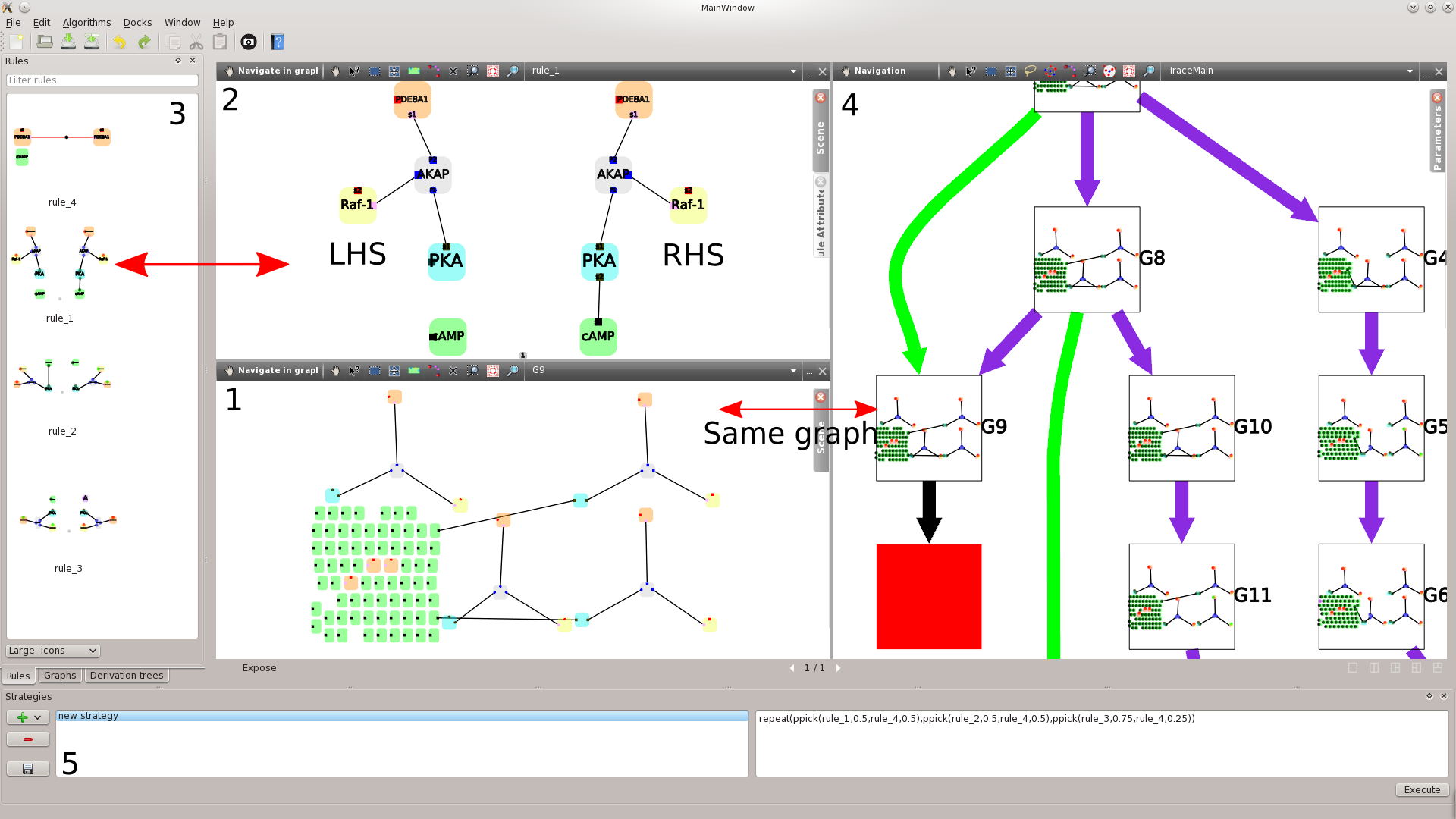}
 \caption{Overview of PORGY: (1) editing one state of the graph being rewritten; (2) editing a rule; (3) all available rewriting rules; (4) portion of the derivation tree, a complete trace of the computing history; (5) the strategy editor.}
\label{fig:overview}
\end{figure}

In this paper we present strategic port graph rewriting as a basis for
the design of PORGY -- a visual, interactive environment for the
specification, debugging, simulation and analysis of complex systems.
PORGY has a graphical interface~\cite{pinaud:hal-00682550} 
and an executable specification language (see Fig.~\ref{fig:overview}), 
where a system is modelled as a port graph together with 
port graph rewriting rules defining its dynamics (Sect.~\ref{PortGraph}).

Reduction strategies define which (sub)expression(s) should be
selected for evaluation and which rule(s) should be applied
(see~\cite{KKK08,BCDK-WRS09} for general definitions). 
Strategies are present
in programming languages such as Clean~\cite{PlasmeijerR:clean},
Curry~\cite{HanusM:curry}, and Haskell~\cite{PeytonJones03} and can be
explicitly defined to rewrite terms in languages such as
{\sc Elan}~\cite{BorovanskyKKMR98}, Stratego~\cite{Vis01.rta},
Maude~\cite{Marti-OlietMV05} or Tom~\cite{TOM-RTA07}.  They are also
present in graph transformation tools such as
PROGRES~\cite{Schurr97b}, AGG~\cite{ErmelRT97},
Fujaba~\cite{NickelNZ00}, GROOVE~\cite{Rensink03},
GrGen~\cite{GeissBGHS06} and GP~\cite{Plump09}.  PORGY's strategy
language draws inspiration from these previous works, but a
distinctive feature 
is that it allows users to
define strategies using not only operators to combine graph
rewriting rules but also operators to define the location in the
target graph where rules should, or should not, apply.

The main contribution of this paper is the definition of a strategic
graph program (Sect.~\ref{Strategy}). It consists of an initial
\emph{located graph} (that is, a port graph with two distinguished
subgraphs $P$ and $Q$ specifying the position where rewriting should
take place, and the subgraph where rewriting is banned, respectively),
and a set of \emph{rewrite rules} describing its dynamic behaviour,
controlled by a \emph{strategy}.  We formalise the concept of strategic graph
program, showing how located graphs generalise the notion of a term
with a rewrite position, and provide a small-step operational
semantics (Sect.~\ref{sect:semantics}) that specifies, for each
strategic graph program, a set of rewrite derivations (i.e., a derivation tree)
generated by  applying the rewrite rules to the initial located graph according 
to the given strategy. 

Strategies are used to control PORGY's rewrite engine: users can
create graph rewriting derivations and specify graph traversals using
the language primitives to select rewriting rules and the position
where the rules apply.  A rewriting position is a subgraph, which 
can be interactively selected (in a visual way), or can be specified
using a \emph{focusing} expression. Alternatively, rewrite
positions could be encoded in the rewrite rules using markers or
conditions~\cite{Plump09}. We prefer to deal with positions directly, 
following Dijkstra's separation of concerns principle~\cite{DijkstraE:selw}.

PORGY and its strategy language were first presented
in~\cite{andrei:2011:inria-00563249:1,FKN:Lopstr}. Unlike those
papers, the notion of port graph considered in this paper includes
attributes for nodes, ports and also edges, which are 
 taken into account in the definition of port graph morphism. In
addition, the strategy language includes a sublanguage to deal with
properties of graphs, which facilitates the specification of rewrite
positions and banned subgraphs (to be protected during rewriting).
Also, in this paper the operational semantics of the language
is formally defined using a transition system that specifies how the derivation
tree is computed for each strategic graph program. In this transition system,
configurations represent the part of the derivation tree that has already been
computed and transitions specify the small-step execution of the commands.
Since the language includes non-deterministic and
probabilistic constructs, the full transition system is probabilistic. 
We give the transition rules for the deterministic sublanguage and
briefly comment on the probabilistic ones.

\section{Port Graph Rewriting}
\label{PortGraph}

Several definitions of graph rewriting are available, using different kinds 
of graphs and rewriting rules (see, for instance,
\cite{Corradini:handbook,HabelMP01,BarendregtHP:tergr,Plump98termgraph,Barthelmann96howto,LafontY:intn}).
In this paper we consider \emph{port graphs} with \emph{attributes} associated to nodes, ports and edges, generalising the notion of port graph introduced in~\cite{AndreiK08c}. 

Intuitively, a port graph is a graph where nodes have explicit
connection points called {\em ports}; edges are attached to
ports. 
Nodes, ports and edges are labelled each one by a  name and attributes.
For instance, a node or a port may have an attribute ``state''
(e.g., with possible values
active/inactive or principal/auxiliary) or attributes defining some properties
such as colour, shape, type, etc.
Attributes may be used to define
the behaviour of the modelled system and for visualisation purposes
(as illustrated in Section~\ref{Examples}).

\paragraph{Port Graph with Attributes.} A {\em labelled port graph with attributes} 
is a tuple $G=( V_G,lv_G,E_G,le_G)$ where:
\begin{itemize}
\item $V_G$ is a finite set of nodes.

\item $lv_G$ is a function that returns, for each $v \in V_G$ with $n$
  ports, a node label $N$ (the node's name), a set
  $\{p_1,\ldots,p_n\}$ of port labels (each with its own set of
  attribute labels and values), and a set of attribute labels (each
  with a value). The node label determines the set of ports and
  attributes. Thus, we may write $\Interface(v) = \Interface(N) = \{p_1,\ldots,p_n\}$.

\item 
$E_G$
is a finite set of edges; each edge has two attachment ports $(v_1,p_1), (v_2,p_2)$, 
where $ v_i\in V_G, p_i\in \Interface(v_i)$. Edges are undirected, 
so  $\langle (v_1,p_1), (v_2,p_2)\rangle$ is an unordered pair, and two nodes 
may be connected by more than one edge on the same ports.
\item $le_G$ is a labelling function for edges, which returns for each
 $e \in E_G$ an edge label,
its attachment ports $(v_1,p_1), (v_2,p_2)$ and its set of attribute labels, 
each with an associated value. 
\end{itemize}

Variables may be used as labels for nodes, ports, attributes and values
in rewrite rules. 

Rewriting is defined using a notion of graph morphism:

\paragraph{Port Graph Morphism.} Let $G$ and $H$ be two port graphs, where $G$ may contain variables but $H$
does not. 
A {\em port graph morphism} $f:G\ra H$ maps nodes, ports, edges with their respective attributes and values from $G$ to  $H$, such that all non-variable 
labels are preserved, the attachment of edges is preserved and the set
of pairs of attributes and values for nodes, ports and edges are also
preserved.  
 If $G$ contains variable labels, the morphism must
instantiate the variables. Intuitively, the morphism identifies a
subgraph of $H$ that is equal to $G$ except for variable occurrences.
For more details we refer the reader to~\cite{CiE2014}.


\paragraph{Port Graph Rewrite Rule.} Port graphs are transformed by applying 
{\em port graph rewrite rules}.  Formally, a port graph rewrite rule  is a port graph consisting of two port graphs 
$L$ and $R$, called the
  \emph{left-hand side} and \emph{right-hand side}, respectively;
 an \emph{arrow} node labelled by $\Ra_n$, where $n$ is the number of ports, and each port in the arrow node has an attribute \textit{type} whose value can be \textit{bridge}, \textit{blackhole} or \textit{wire}; and
a set of edges that each connect a port of the arrow node to ports
 in $L$ or $R$. 
This set of edges must satisfy the following conditions:
\begin{enumerate}
\item
A port of type bridge must have edges connecting it to $L$ and to $R$ (one edge to $L$ and
 one or more to $R$).
\item
A port of type blackhole must have edges connecting it only to $L$ (at least one edge).
\item
A port of type wire must have exactly two incident edges from $L$ and no edges connecting 
it to $R$.
\end{enumerate}
The arrow node and arrow-edges are omitted if
they are obvious from $L$ and $R$.

The left-hand side of the rule, also
called pattern, is used to identify subgraphs in a given graph, which
are then replaced by the right-hand side of the rule. The arrow node 
describes the way the new subgraph should be linked to
the remaining part of the graph, to avoid dangling
edges~\cite{HabelMP01,Corradini:handbook} during rewriting.

\paragraph{Derivation.} A port graph $G$ {\em
  rewrites} to $G'$ using the rule $r= L \Ra R$ and a morphism $g$
from $L$ to $G$, written $G \ra_r^g G'$, if $G'$ is obtained from $G$
by replacing $g(L)$ by $g(R)$ in $G$ and connecting $g(R)$ to the rest of $G$
as specified by $r$'s arrow node. We write $G\ra_{\R}G'$ if $G \ra_r^g G'$ using $r
\in \R$.  This induces a reflexive and transitive relation on port
graphs, called \emph{the rewriting relation}, denoted by $\ra_{\R}^*$.
Each {\em rule application} is a rewriting step and a {\em
  derivation}, or computation, is a sequence of rewriting steps.

\paragraph{Derivation Tree.} Given  a port graph $G$ and a set of port graph rewrite rules $\R$, the 
\emph{derivation tree} of $G$, written $DT(G,\R)$, is a labelled tree such that 
the root is  labelled by  the initial port graph $G$,  and its children are the 
roots of  the
derivation trees 
 $DT(G_i,\R)$ such that  $G\ra_{\R}G_i$.
The edges of the derivation tree are labelled with the rewrite rule
and the morphism used in the corresponding rewrite step.
We will use \emph{strategies} to specify the rewrite derivations of interest.

\section{Strategic graph programs}
\label{Strategy}

\begin{definition}[Located graph.]
\label{def:locatedgraph}
A \emph{located graph} $G_{P}^{Q}$ consists of a port graph $G$ and two
distinguished subgraphs $P$ and $Q$ of $G$, called respectively the \emph{position
  subgraph}, or simply \emph{position}, and the \emph{banned subgraph}. 
\end{definition}

In a located graph $G_{P}^Q$, $P$ represents the subgraph of $G$ where
rewriting steps may take place (\emph{i.e.}, $P$ is the focus of the
rewriting) and $Q$ represents the subgraph of $G$ where rewriting
steps are forbidden.  We give a precise definition below;
the intuition is that subgraphs of $G$ that overlap with $P$ may be
rewritten, if they are outside $Q$.  The subgraph $P$ generalises the notion of 
rewrite position in a term: if $G$ is the tree representation of a term $t$ 
then we recover the usual notion of rewrite position $p$ in $t$ by setting $P$
to be the node at position $p$ in the tree $G$, and $Q$ to be the part of the
tree above $P$ (to force the rewriting step to apply at $p$, \emph{i.e.}, downwards from
the node $P$).

When applying a port graph rewrite rule, not only the underlying graph
$G$ but also the position and banned subgraphs may change.  A
\emph{located rewrite rule}, defined below, specifies two
disjoint subgraphs $M$ and $N$ of the right-hand side $R$ that are
used to update the position and banned subgraphs, respectively.  If
$M$ (resp.\ $N$) is not specified, $R$ (resp.\ the empty graph
$\emptyset$) is used as default.  Below, we use the operators $\cup, \cap,
\setminus$ to denote union, intersection and complement of port graphs. These operators
are defined in the natural way on port graphs considered as sets of nodes, ports
and edges.

\begin{definition}[Located rewrite rule.]
\label{def:locrew}
A \emph{located rewrite rule} is given by a port graph
rewrite rule $L \Ra R$, and optionally a subgraph $W$ of $L$ and two disjoint subgraphs $M$ and $N$ of $R$.
It is denoted $L_W\Ra R_{M}^N$. We write $G_P^Q \ra_{L_W \Ra R_{M}^N}^{g} {G'}_{P'}^{Q'}$ and say that the
located graph $G_P^Q$ \emph{rewrites to} ${G'}_{P'}^{Q'}$ \emph{using} $L_W \Ra R_{M}^N$ \emph{at position} $P$
\emph{avoiding} $Q$, if $G\ra_{L \Ra R}G'$
with a morphism $g$ such that $g(L)\cap P = g(W)$ or  simply $g(L)\cap P\neq \emptyset$
if $W$ is not provided, and 
$g(L)\cap Q = \emptyset$. The new position subgraph $P'$ and banned subgraph $Q'$ in $G'$
are defined as $P'=(P \setminus g(L)) \cup g(M)$, $Q'= Q \cup g(N)$; if $M$ (resp.\ $N$) are not
provided then we assume $M = R$ (resp.\ $N = \emptyset$).
\end{definition}

In general, for a given located rule $L_W \Ra R_{M}^N$ and located graph
$G_P^Q$, more than one morphism $g$, such that $g(L)\cap P = g(W)$ 
and $g(L)\cap Q=\emptyset$, may exist (\emph{i.e.}, several rewriting
steps at $P$ avoiding $Q$ may be possible). Thus, the application of
the rule at $P$ avoiding $Q$ produces a \emph{set of
  located graphs}.

To control the application of rewriting rules, we introduce a strategy
language whose syntax is shown in Table~\ref{tab:syntax-strategies}.
\emph{Strategy expressions} are generated by the grammar rules from
the non-terminal $S$. A strategy expression combines
\emph{applications} of located rewrite rules, generated by the
non-terminal $A$, and \emph{position updates}, generated by the
non-terminal $U$ with \emph{focusing expressions} generated by $F$.
The application constructs and some of the strategy constructs are
strongly inspired by term rewriting languages such as {\sc
  Elan}~\cite{BorovanskyKKMR98}, Stratego~\cite{Vis01.rta} and
Tom~\cite{TOM-RTA07}.  Focusing operators are not present in term
rewriting languages where the implicit assumption is that the rewrite
position is defined by traversing the term from the root downwards.

The syntax presented here extends the one in~\cite{FKN:Lopstr} by
including a language to define subgraphs of a given graph by selecting nodes
that satisfy some simple properties (see Table~\ref{tab:syntax-property}).

\begin{table}[!t]
\centering
\fbox{
\renewcommand{\arraystretch}{1.5}
\begin{tabular}{l}
  Let $L, R$ be port graphs; $M,N$ positions;  $n\in \mathbb{N}$; $\pi_{i=1 \ldots n} \in [0,1]$; $\sum\limits_{i=1}^{n}\pi_i = 1$ 
  \\
\begin{tabular}{rrrl}
{\bf (Strategies)} & $S$ & $::=$ & $   A \mid U \mid S ; S \mid \Repeat{S} \mid \whiledoo{S}{S} $ \\  
& & $\mid$ & $\orelse{S}{S}\mid \ifthenelsee{S}{S}{S} $\\
& & $\mid$ & $ \ppick{S_1,\pi_1,\dots ,S_n,\pi_n}  $\\
 {\bf (Applications)} & $A$ & $::=$ & $\Id \mid \Fail \mid \all{T} \mid \one{T} $ \\
 {\bf (Transformations)} & $T$ & $::=$ & $L_W \Ra R_{M}^{N}$ \\ 
 {\bf (Position Update)} & $U$ & $::=$ & $ \setPos{F} \mid \setBan{F}\mid \isEmpty{F} $\\
 {\bf (Focusing)} &  $F$ & $::=$ & $\CrtGraph \mid  \CrtPos \mid \CrtBan \mid \AllNgb{F}$\\
  & & $ \mid$ & $ \OneNgb{F} \mid \NextNgb{F} \mid \Property{\rho}{F}$\\
  & & $\mid$ &  $F \cup F \mid 
   F \cap F \mid  F \setminus F\mid \emptyset $
\end{tabular}
\end{tabular}
}
\caption{Syntax of the strategy language.}\label{tab:syntax-strategies}
\end{table}

\begin{table}[!t]
\centering
\fbox{
\renewcommand{\arraystretch}{1.5}
\begin{tabular}{l}
Let \textit{attribute} be an attribute label; $a$ a valid value for the given attribute label;\\ 
\textit{function-name} the name of a built-in or user-defined function.
 \\
\begin{tabular}{rrrl}
 {\bf (Properties)} & $\rho$ & $:=$ &  $(Elem, Expr) | ({\tt Function},~\textit{function-name})$\\
& $Elem$ & $:=$ & ${\tt Node} \mid {\tt Edge} \mid {\tt Port}$\\
 & $Expr$& $:=$ & 
${\tt Label} == a  \mid {\tt Label} ~ !\!\! = a  \mid \textit{attribute} ~ Relop ~\textit{attribute}$ \\
& & & $ \mid \textit{attribute} ~ Relop ~a$  \\
& $Relop$ & $:=$ & $ == ~ \mid ~  !\!\! = ~ \mid ~ > ~ \mid ~ < ~ \mid ~ >= ~ \mid ~ <= $
\end{tabular}
\end{tabular}
}
\caption{Syntax of the Property Language.}\label{tab:syntax-property}
\end{table}

The focusing constructs are a distinctive feature of our language.
They are used to define positions for rewriting in a graph, or to
define positions where rewriting is not allowed.  They denote
functions used in strategy expressions to change the positions $P$ and
$Q$ in the current located graph (e.g., to specify graph
traversals). We describe them briefly below.

\begin{itemize}
\item $\CrtGraph$, $\CrtPos$ and $\CrtBan$, applied to a located graph
  $G_P^Q$, return respectively $G$, $P$ and $Q$.

\item $\tt{AllNgb}$, $\tt{OneNgb}$ and $\tt{NextNgb}$ denote functions
  that apply to pairs consisting of a located graph $G_P^Q$ and a
  subgraph $G'$ of $G$. If $Pos$ is an expression denoting a subgraph
  $G'$ of the current graph $G$, then $\AllNgb{Pos}$ is the subgraph
  of $G$ consisting of all immediate successors of the nodes in $G'$,
  where an immediate successor of a node $v$ is a node that has a port
  connected to a port of $v$. $\OneNgb{Pos}$ returns a subgraph of $G$
  consisting of one randomly chosen node which is an immediate
  successor of a node in $G'$. $\NextNgb{Pos}$ computes all successors
  of nodes in $G'$ using for each node only the port labelled ``next''
  (so $\NextNgb{Pos}$ returns a subset of the nodes returned by
  $\AllNgb{Pos}$).

\item
$\Property{\rho}{F}$ is used to select a subgraph of a given graph, satisfying a certain property, specified by $\rho$. It can be seen as a filtering construct: if the focusing 
expression $F$ 
generates a subgraph $G'$ then $\Property{\rho}{F}$ returns a
  subgraph containing only the nodes and edges from $G'$ that satisfy the
  decidable property $\rho$.
It typically tests a property on nodes, ports, or edges,  allowing us for instance to
   select the subgraph of  nodes with active ports: $\Property{(\texttt{Port}, State == active)}{F}$.
It is also possible to specify a function to be used to compute the subgraph: $\Property{(\texttt{Function},Root)}{\CrtGraph}$ uses the built-in (or user-defined) function $Root$ to compute a specific subgraph from the current graph. 
  
\item $\cup$, $\cap$ and $\setminus$ are union, intersection and complement of port graphs which may be used to combine multiple \textsf{Property} operators; $\emptyset$ denotes the empty graph.
\end{itemize}

Other operators can be derived from the language constructs. A useful
example is the  {\tt not} construct:
 \begin{itemize}
   \item $\nott{S}\triangleq \ifthenelsee{S}{\Fail}{\Id}$. It fails if $S$ succeeds and succeeds if $S$ fails.
\end{itemize}

\begin{definition}[Strategic graph program.]
\label{def:graphprog}
A \emph{strategic graph program} consists of a finite set of located rewrite
rules ${\cal R}$, a strategy expression $S$ (built with $\cal R$ using
the grammar in Table~\ref{tab:syntax-strategies}) and a located graph $G_{P}^{Q}$. 
We denote it
$\sema{S_{\cal R}, G_P^Q}$, or simply $\sema{S, G_P^Q}$ when ${\cal
  R}$ is clear from the context.
\end{definition}

\section{Semantics of strategic graph programs}
\label{sect:semantics}

Intuitively, a strategic program consists of an initial port graph,
together with a set of rules that will be used to reduce it, following
the given strategy.  Formally, the semantics of a strategic graph
program $\sema{S, G_P^Q}$ is specified using a transition system (that
is, a set of configurations with a binary relation on configurations),
defining a \emph{small step} operational semantics in the style
of~\cite{PlotkinSOS}.

\begin{definition}[Configuration.]
A \emph{configuration} is a multiset $\{O_1 , \dots , O_n\}$ where each $O_i$ is
 a strategic graph program.
\end{definition}

 Given a strategic graph program $\sema{S_{\cal R}, G_P^Q}$, we will define sequences of
transitions according to the strategy $S$, starting from  the \emph{initial configuration} 
$\{\sema{S, G_P^Q}\}$. 
A configuration is \emph{terminal} if no transitions can be performed. 

We will prove that all terminal configurations in our transition
system consist of \emph{results},
denoted by $V$, of the form $[\Id , G_P^Q]$ or $[\Fail , G_P^Q]$.   
In other words, there are no blocked programs: the transition system
ensures that, for any configuration, either there are transitions to
perform, or we have reached results. 

Below we provide the transition rules for the core sublanguage, that
is, the sublanguage that does not include the non-deterministic operators
\one{}, \orelse{}{}, \ppick{}, \Repeat{} and \OneNgb{}. 

\begin{definition}[Transitions]
\label{def:trans}
The transition relation $\lra$ is a binary relation on configurations, defined as follows:
$$\{O_1,\ldots,O_k,V_1,\ldots,V_j\} \lra \{O'_{11},\ldots,O'_{1m_1},\ldots,O'_{km_k},V_1,\ldots, V_j\}$$
if $O_i \ra \{O'_{i1},\ldots,O'_{im_i}\}$, for  $1 \leq i \leq k$,
where $k\geq 1$ and where some of
the $O'_{ij}$ might be results.

The auxiliary relation $\ra$ is defined below using axioms and rules.
\end{definition}

A configuration $\{O_1,\ldots,O_k,V_1,\ldots,V_j\}$ is a multiset of
graph programs, representing a partially computed derivation
tree. Each element in the configuration represents a node in the
derivation tree associated to the initial graph program. Some of the
elements may already be results. The transition relation performs
reductions in parallel at all the positions in the derivation tree
where there is a reducible graph program $O_i$.

\begin{definition} 
\label{def:aux-trans} 
The transition relation $\ra$ on individual strategic graph programs is defined by induction.

There are no axioms/rules defining transitions for a program where the strategy is $\Id$ or $\Fail$ (these are terminal).

\textit{Axioms for the operator {\sf all}:}
\[\begin{prooftree} 
\justifies
[\all{L_W \Ra R_M^N},{G_P^Q}] \ra  \{[\Id,{{G_1}_{P_1}^{Q_1}}],\dots,[\Id,{{G_k}_{P_k}^{Q_k}}]\}
\using{LS_{L_W \Ra R_M^N}({G_P^Q})=\{{G_1}_{P_1}^{Q_1},\ldots,{G_k}_{P_k}^{Q_k}\} }
\end{prooftree}
\]
\[\begin{prooftree}
\justifies 
[\all{L_W \Ra R_M^N},{G_P^Q}] \ra \{ [\Fail,{G_P^Q}] \}
\using{ {LS_{L_W \Ra R_M^N}({G_P^Q})=\emptyset }}
\end{prooftree}\] 
where $LS_{L_W \Ra R_M^N}({G_P^Q})$, the \emph{set of legal reducts of $G_P^Q$ for  ${L_W \Ra R_M^N}$}, or \emph{legal set} for short, contains all the located graphs
${G_i}_{P_i}^{Q_i} $ ($1 \leq i \leq k$) such that $G_P^Q \ra_{L_W \Ra R_M^N}^{g_i} {G_i}_{P_i}^{Q_i}$ and  $ g_1,\ldots,g_k$ are  pairwise different.

As the name of the operator indicates, all possible applications of the rule are considered in one step. The strategy fails if the rule is not applicable.

\textit{Position Update and Focusing.}
Next we give the semantics of the commands that are used to specify and update positions \emph{via}
focusing constructs. 
The focusing expressions  generated by the grammar for the non terminal 
$F$ in Tab.~\ref{tab:syntax-strategies}  have a functional semantics (see below). 
In other words, an expression $F$ denotes a function that applies to the current located graph, and computes a subgraph of $G$.
Since there is no ambiguity, the function  denoted by the expression
$F$ is also called $F$. 
\[
\begin{prooftree}
\justifies
[\setPos{F} , {G_P^Q} ]  \ra \{[\Id , {G_{P'}^Q}] \}  
\using{ F(G_P^Q) = P'}
\end{prooftree}
\qquad
 \begin{prooftree}
 \justifies
 [\setBan{F} , {G_P^Q}]  \ra  \{[\Id , {G_{P}^{ Q'}}] \}
 \using{ F(G_P^Q) = Q'}
 \end{prooftree}
 \]
\[
\begin{prooftree}
\justifies
[\isEmpty{F}, {G_P^Q}]  \ra \{ [\Id , {G_{P}^{Q}}]\}
\using{F(G_P^Q) = \emptyset}
\end{prooftree}
\]
\[
\begin{prooftree}
\justifies
[\isEmpty{F}, {G_P^Q}]  \ra  \{[\Fail , {G_{P}^{Q}}] \}
\using{F(G_P^Q) \neq \emptyset}
\end{prooftree}\]

\[\begin{array}{lclclclclcl}
\CrtGraph(G_P^Q) &= & G & \qquad &  \CrtPos(G_P^Q) &= &P &\qquad & 
\CrtBan(G_P^Q) &= &Q\\
\AllNgb{F}(G_P^Q) &= &G'& \qquad &\multicolumn{7}{l}{\mbox{where $G'$ consists of all 
immediate successors of}}\\
 & & & &\multicolumn{7}{l}{\mbox{nodes in  $F(G_P^Q)$}}\\ 
\NextNgb{F}(G_P^Q) &= &G'& \qquad &\multicolumn{7}{l}{\mbox{where $G'$ consists of 
the immediate successors,}}\\
& & & &\multicolumn{7}{l}{\mbox{\emph{via} ports labelled ``next'', of nodes in  $F(G_P^Q)$}}\\
\Property{\rho}{F}(G_P^Q) &= &G'& \qquad &\multicolumn{7}{l}{\mbox{where $G'$ consists of all nodes in $F(G_P^Q)$ satisfying $\rho$}} \\
(F_1~ op ~F_2) (G_P^Q) &= & \multicolumn{7}{l}{F_1(G_P^Q)~ op~ F_2(G_P^Q) 
\mbox{~~where $op$ is $\cup,\cap,\setminus$}}
\end{array}\]

Note that with the semantics given above for $\setPos{}$ and $\setBan{}$, it is possible for 
$P$ and $Q$ to have a non-empty intersection.  A rewrite rule can still 
apply if the redex overlaps  $P$ but not $Q$. 

\textit{Sequence.}
The semantics of sequential application, written $S_1; S_2$, is defined by  two axioms and a rule: 
\[\begin{prooftree}
\justifies
 [\Id;S,G_P^Q]   \ra \{ [S,G_P^Q]\}
\end{prooftree}
\quad
\begin{prooftree}
\justifies
 [\Fail;S,G_P^Q]   \ra \{ [\Fail,G_P^Q]\}
\end{prooftree}
\]
\[\begin{prooftree}
[S_1, {G_P^Q}]\ra \{[S_1^1,{G_1}_{P_1}^{Q_1}],\dots,[S_1^k,{G_k}_{P_k}^{Q_k}]\} 
\justifies
[S_1 ; S_2, {G_P^Q}] \ra  \{[S_1^1;S_2,{G_1}_{P_1}^{Q_1}],\dots,[S_1^k;S_2,{G_k}_{P_k}^{Q_k}]\} 
\end{prooftree}
\]

The rule for sequences ensures that $S_1$ is applied first.

\textit{Conditional.}
The behaviour of the strategy $\ifthenelsee{S_1}{S_2}{S_3}$ depends on the result of the strategy $S_1$. If $S_1$ succeeds on (a copy of) the current located graph, then $S_2$ is applied to the current graph, otherwise $S_3$ is applied.  
\[\begin{prooftree}
 \{[S_1,G_P^Q]\}\lra^* M ~ s.t. ~ [\Id,G']\in M
\justifies
[\ifthenelsee{S_1}{S_2}{S_3} ,{G_P^Q} ]  \ra\{ [S_2 ,{G_P^Q}]\}
\end{prooftree}
\]
\[
\begin{prooftree}
 \{[S_1,G_P^Q]\} \lra^* \{[\Fail,G_1], \ldots,[\Fail,G_n]\} 
\justifies
[\ifthenelsee{S_1}{S_2}{S_3} ,{G_P^Q} ]  \ra \{[S_3 ,{G_P^Q}]\}
\end{prooftree}
\]

\textit{While loop.}
 Iteration is defined using a conditional as follows: 
\[\begin{prooftree}
\justifies
[ \whiledoo{S_1}{S_2} ,{G_P^Q} ]   \ra 
\{[\ifthenelsee{S_1}{S_2;\whiledoo{S_1}{S_2}}{\Id}, {G_P^Q}]\}
\end{prooftree}\]
\end{definition}

Note that $S_1$ used as a condition in the two constructs above may produce 
some successes and some failure results. Also, in general the strategy $S_1$ 
could be non-deterministic and/or non-terminating. To avoid non-deterministic 
conditions in \textsf{if} and  \textsf{while commands}, 
 the class $Cond$ of strategies generated by the following grammar should be used:

$$Cond :: = Cond; Cond \mid \Id \mid \Fail \mid \all{T} \mid \isEmpty{F} \mid \nott{Cond}$$
where $F$ should also be deterministic:
$$F ::= \AllNgb{F} \mid \NextNgb{F} \mid \Property {\rho}{F} \mid \cup\mid \cap \mid \setminus\mid \emptyset$$

However, using non-deterministic constructs in the condition is not necessarily unsafe:
if $R$ is a located rule, we could, for instance, write $\ifthenelsee{\one{R}}{S_2}{S_3}$
to perform either $S_2$ or $S_3$, depending on whether $R$ is applicable at the current position or not.
Also note that although  the strategy $\one{R}$ is non-deterministic, the strategy $\nott{\one{R}}$ is deterministic (we are simply testing whether $R$ can be applied or not).

We finish this section by giving the intuition for the semantics of the remaining constructs. 

To define the semantics of the non-deterministic and probabilistic
constructs in the language, we generalise the transition relation.
Let us denote by $\ra_\pi$ a transition step with probability $\pi$. The
relation $\ra$ defined above  can be seen as a
particular case where $\pi =1$, that is, $\ra$ corresponds to $\ra_1$.
The relation $\lra$ on configurations also becomes probabilistic: 
$$\{O_1,\ldots,O_k,V_1,\ldots,V_j\} \lra_\pi \{O'_{11},\ldots,O'_{1m_1},\ldots,O'_{km_k},V_1,\ldots,V_j\}$$
if $O_i \ra_{\pi_i} \{O'_{i1},\ldots,O'_{im_i}\}$, for  $1 \leq i \leq k$ (where $k\geq 1$ and some of
the $O'_{ij}$ might be results) and $\pi = \pi_1 \times \cdots \times \pi_k$. 

We write $M \lra^*_\pi M'$ if there is a sequence of transitions
$\lra_{\pi_i}$ from configuration $M$ to $M'$, such that the product of
probabilities is $\pi$.  

We can define transition rules for the remaining 
constructs in the strategy language as follows.

\textit{Probabilistic Choice of Strategy: }
{
\[\begin{prooftree}
\justifies
[\ppick{S_1,\pi_1,\ldots ,S_n,\pi_n} , {G_P^Q}] \ra_{\pi_j}  \{[S_j,{G_P^Q}]\}
\end{prooftree}\]
}

\textit{Non-deterministic Choice of Reduct:}
The non-deterministic $\one{}$ operator takes as argument a
  rule. 
  It randomly selects only one amongst the set of legal reducts $LS_{L_W \Ra R_M^N}({G_P^Q})$. Since all of them have the same probability of
being selected, in the axiom below $\pi = 1/|LS_{L_W \Ra R_M^N}({G_P^Q})|$.

\[
\begin{prooftree}
\justifies
[\one{L_W \Ra R_M^N},{G_P^Q}] \ra_\pi  \{[\Id,{{G'}_{P'}^{Q'}}]\} 
\using{ {{G'}_{P'}^{Q'}} \in LS_{L_W \Ra R_M^N}({G_P^Q})}
 \end{prooftree}
\]
\[
\begin{prooftree}
\justifies
[\one{L_W \Ra R_M^N},{G_P^Q}]  \ra_1 \{ [\Fail,{G_P^Q}] \}
\using{ LS_{L_W \Ra R_M^N}({G_P^Q})= \emptyset} 
 \end{prooftree}\]

We omit the rules for $\textsf{orelse}$ and $\textsf{repeat}$, and for the
commands $\setPos{F}$, $\setBan{F}$ and $\isEmpty{F}$, which are non-deterministic if
the expression $F$ is non-deterministic.  Note that in focusing
constructs, non-determinism is introduced by the operator $\OneNgb{F}$.


\section{Examples} 
\label{Examples}

Using focusing (specifically the {\tt Property} construct), we can
create concise strategies that perform traversals\footnote{Working
  examples can be downloaded from
  \url{http://tulip.labri.fr/TulipDrupal/?q=porgy}.}. In this way, we
can for instance switch between \emph{outermost and innermost term rewriting}
(on trees). This is standard in term-based languages such as {\sc
  Elan}~\cite{BorovanskyKKMR98} or
Stratego~\cite{Vis01.rta, BKVV08}; here we can also define
traversals in graphs that are not trees. More examples can be found
in~\cite{andrei:2011:inria-00563249:1, pinaud:hal-00682550, FKN:Lopstr}.

The following strategy allows us to check if a graph is connected 
using a standard connectivity test. Assuming that all nodes of the initial 
graph have the Boolean attribute \emph{state} set to false, we just need 
one rewriting rule, which simply sets \emph{state} to $true$ on a node. 
We start with the strategy \emph{pick-one-node} to randomly select a node $v$ as a starting point. Then, the rule is applied to all neighbours of $v$. When the rule cannot be applied any longer, the position subgraph is set to all neighbours of the previously used nodes which still have \emph{state} set to $false$ (\emph{visit-neighbours-at-any-distance}). The strategy continues until the position subgraph is empty. If the rule can still be applied somewhere in the graph,  there is a failure (\emph{check-all-nodes-visited}). Note the use of attributes and focusing constructs to traverse the graph. Below the strategy $R$ is an abbreviation for $\one{R}$.
\begin{align*}
\textit{pick-one-node: }&\setPos{\CrtGraph};\\
&R;\\
&\setPos{\Property{(Node, state==true)}{\CrtGraph}};\\
\textit{visit-neighbours-at-any-distance: }&\setPos{\AllNgb{\CrtPos}};\\
&\whiledoo{\nott{\isEmpty{\CrtPos}}}{\\
&~~~\ifthenelsee{R} {R} {\\
&~~~\setPos{\AllNgb{\CrtPos} \setminus \\
&\Property{(Node, state==true)}{CrtGraph}}}};\\
\textit{check-all-nodes-visited: }&\setPos{\CrtGraph};\\
& \nott{R}
\end{align*}

\begin{figure}[ht]
\centering
\includegraphics[width=0.8\columnwidth, keepaspectratio]{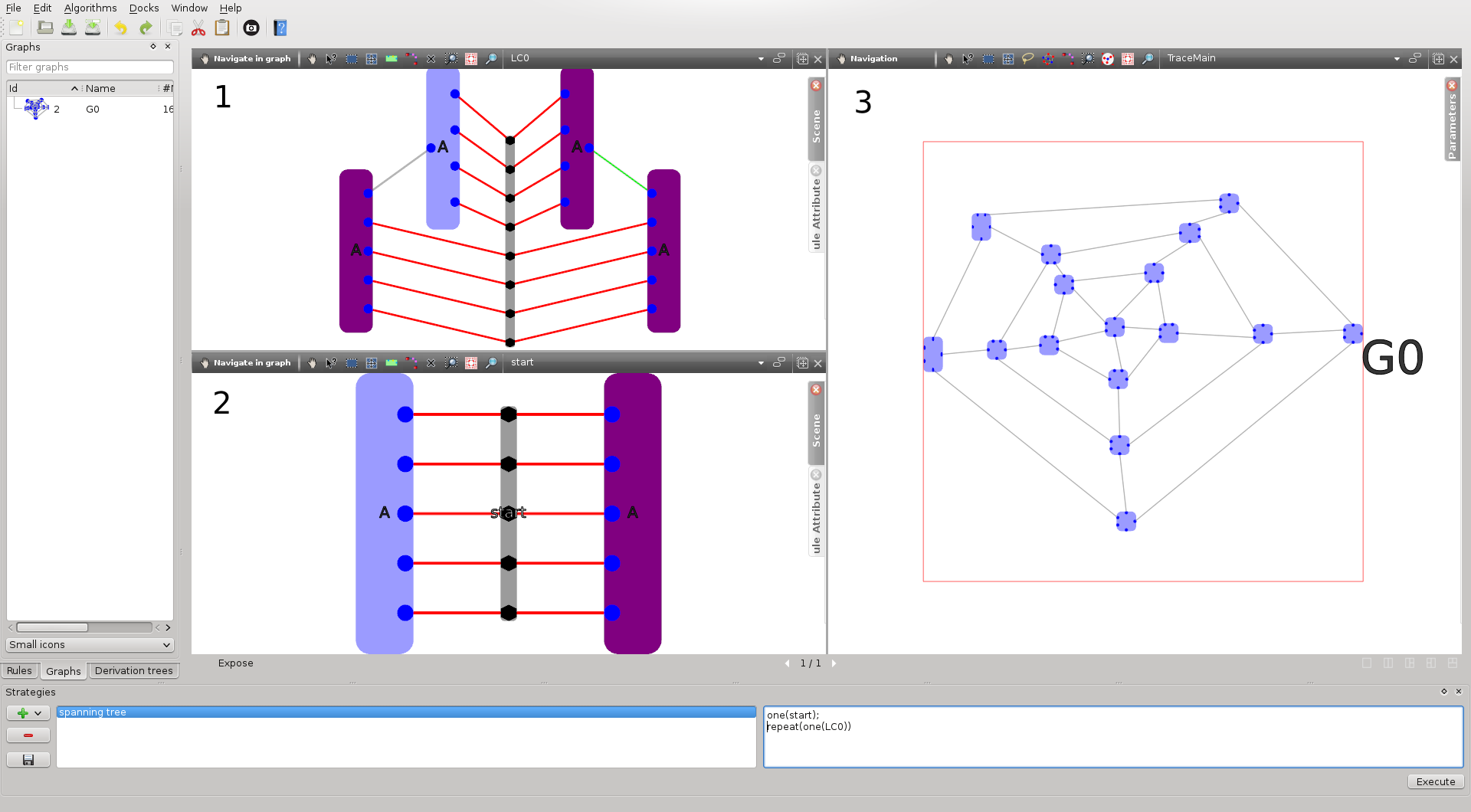}
 \caption{Computation of a spanning tree. Panel 1 shows the rule $LC0$. It is used to add nodes and edges to the spanning tree. Panel 2  shows the rule which sets the root of the tree. Panel 3 is the root of the derivation tree with the graph used for computation.}
\label{fig:spanning}
\end{figure}

The next example uses node and edge attributes encoded inside two rules to build a spanning tree from a graph (see Fig.~\ref{fig:spanning}). The rules are: $start$, which is used to select the root of the tree, and $LC0$, which builds a branch of the tree. $LC0$ works as follows: given an existing node $v$ of the tree, if $v$ is linked to another node not part of the tree with an edge also not part of the tree, add both of them to the tree. The strategy used to build one spanning tree is very simple:

\begin{align*}
 \one{start};\\
 \Repeat{\one{LC0}}
\end{align*}

\begin{figure}[ht]
\centering
\setlength{\fboxsep}{0pt}%
\setlength{\fboxrule}{1pt}%
\fbox{\includegraphics[width=0.9\columnwidth, keepaspectratio]{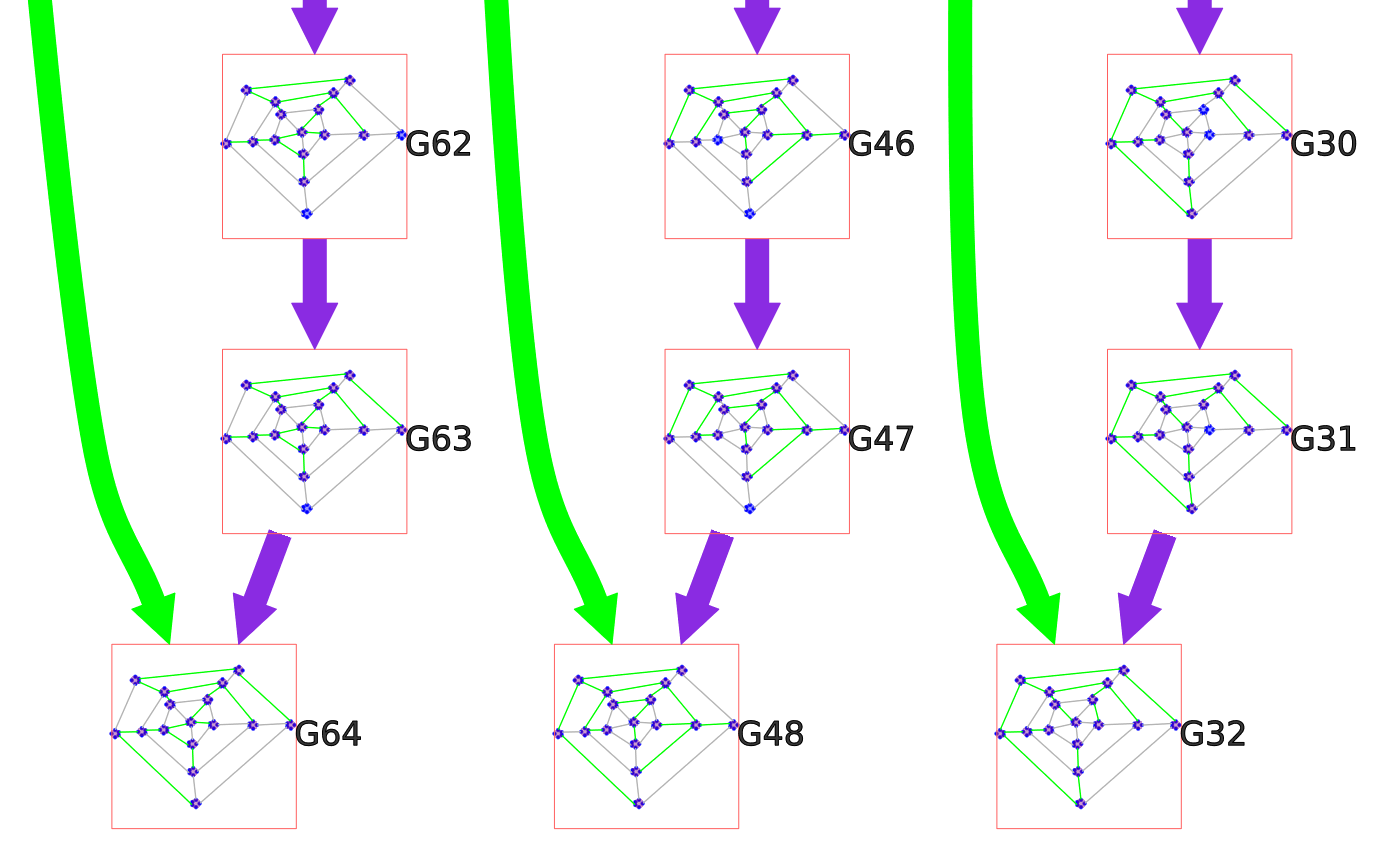}}
 \caption{Three spanning trees given by applying the strategy three times from $G0$ (see Fig.~\ref{fig:spanning}).}
\label{fig:derivation_tree_spanning_tree}
\end{figure}

If one wants all possible spanning trees, \one{} has simply to be replaced by \all{}. Figure~\ref{fig:derivation_tree_spanning_tree} shows the results for three applications of the strategy.

\section{Properties}
\label{Properties}

In this section we discuss termination and completeness of the
strategy language.  

\begin{definition}[Termination.]
A strategic graph program $\sema{S,G_P^Q}$ is \emph{terminating}
 if there is no infinite
transition sequence from the initial configuration $\{\sema{S,
  G_P^Q}\}$.  It is \emph{weakly
  terminating} if a configuration having at least one
result 
can be reached.
\end{definition}

\begin{definition}[Result set.]
The \emph{result set}
associated to a sequence of transitions out of the 
configuration $\{\sema{S, G_P^Q}\}$ is the set of all the
results in the configurations in the sequence.
Given a strategic graph program $\sema{S , G_P^Q}$, if the sequence of transitions out of the initial configuration
$\{\sema{S, G_P^Q}\}$ ends in a terminal configuration then the
 result set of the sequence is a \emph{complete result set} for the program
$\sema{S, G_P^Q}$.  
If a strategic
graph program does not reach a terminal configuration (in case of
non-termination) then the complete result set is undefined ($\bot$).
\end{definition}

Note that there may exist more than one sequence of transitions out of the initial
configuration $\{\sema{S, G_P^Q}\}$ ending in a terminal
configuration. However, for the core part of the language (that is, excluding the
non-deterministic constructs $\ppick{}$, $\orelse{}{}$, $\Repeat{}$, $\one{}$, and $\OneNgb{}$), strategic graph programs have at most one terminal
configuration (none if the program is non-terminating). As a consequence, 
each strategic graph program in the core language 
has at most one complete result set (Prop.~\ref{prop:res-set}).

Graph programs are not terminating in general, however we can identify
a terminating sublanguage (\emph{i.e.}\  a sublanguage for which the transition
relation is terminating). We can also characterise the terminal configurations.
The next lemma is useful for the termination proof:

\begin{lemma}
\label{lem:seq}
If $[S_1,G_P^Q]$ is  terminating and $S_2$ is such that
$[S_2,{G'}_P^Q]$ is  terminating for any ${G'}_P^Q$, then
$[S_1;S_2,G_P^Q]$ is  terminating.
\end{lemma}

\begin{property}[Termination]
The sublanguage that excludes the $\while{}{}$ and $\Repeat{}$
constructs is  terminating.
\end{property}

\begin{property}[Progress: Characterisation of Terminal Configurations]
\label{normalformproperty}
For every strategic graph program $[S,G_P^Q]$ that  is not a result
(\emph{i.e.}, $S \neq \Id$ and $S\neq \Fail$), there exists a configuration $C$ such that $\{[S,G_P^Q]\} \ra C$.
\end{property}

\begin{proof}
By induction on $S$. According to the definition of transition in Sect.~\ref{sect:semantics}, for every
strategic graph program $[S,G_P^Q]$ different from $[\Id,G_P^Q]$ or
$[\Fail,G_P^Q]$, there is an axiom or rule that applies (it suffices to check all the cases in the
grammar for $S$).
\end{proof}

The language contains non-deterministic operators in each of its syntactic categories: $\OneNgb{}$ for Position Update, 
$\one{}$ for Applications and $\ppick{}$, $\orelse{}{}$ and $\Repeat{}$ for Strategies. For the sublanguage that excludes them, we have the
property:

\begin{property}[Unique Complete Result Set]
\label{prop:res-set}
Each strategic graph program in the sublanguage that excludes $\OneNgb{}$, $\one{}$, $\ppick{}$, $\orelse{}{}$ and $\Repeat{}$ has at most one \emph{complete result set}.
\end{property}

\begin{proof}
If we exclude those constructs, the transition system is deterministic, so there is at most one sequence of transitions out of any initial configuration. Hence there is at most one terminal configuration and therefore at most one complete result set.
\end{proof}

With respect to the computation power of the language, it is easy to
state the Turing completeness property. The proof is similar to
that in~\cite{HabelA:comp}.

\begin{property}[Turing Completeness]
The set of all strategic graph programs $\sema{S_{\cal R}, G_P^Q}$  is Turing
complete, \emph{i.e.}\  can simulate any Turing machine.
\end{property}

\section{Implementation}\label{sect:implementation}

PORGY is implemented on top of the visualisation framework Tulip~\cite{auber:2012:hal-00659880:1} as a set of Tulip plugins. The strategy language is one of these plugins. A version of Tulip bundled with PORGY can be downloaded from \url{http://tulip.labri.fr/TulipDrupal/?q=porgy}. 

 Our first challenge was to implement port graphs, because Tulip only supports nodes and edges from a graph theory point of view.
We had to develop an abstract layer on top of the Tulip graph library to be able to easily work with port graphs. 

When applying a rule $L \Rightarrow R$ on a graph $G$, the first operation is to compute the morphism between the left-hand side $L$ and $G$. This problem, known as the graph-subgraph isomorphism, still receives great attention from the community. We have implemented Ullman's original algorithm~\cite{UllmanJ:subgip} because its implementation is straightforward and it is  used as a reference in many papers.

The derivation tree is implemented with the help of metanodes (a node which represents a graph) and quotient graph functionalities of Tulip (a graph of metanodes). Each node of the derivation tree represents a graph $G$, except red nodes which represent failures ($\Fail$). Inside each node, the user sees an interactive drawing of the graph (see panel 4 of Fig.~\ref{fig:overview}). 
See~\cite{pinaud:hal-00682550} for more details about the interactive features of PORGY and how we implemented them.

The strategy plugin is developed with the Spirit C++ library from Boost\footnote{see \url{http://www.boost.org/libs/spirit} for more details}. This plugin works as a compiler: its inputs are a strategy defined as a text string and the Tulip graph datastructure, the output are low-level Tulip graph operations. Boost (precisely its Random library) is also used to generate the random numbers needed for the probabilistic operators. For instance, we use a non-uniform generator for \ppick{} to be able to choose a strategy following the given probabilities.

\section{Conclusion}
\label{Conclusion}
The strategy language defined in this paper is part of PORGY, an environment for visual modelling and analysis of
complex systems through port graphs and port graph rewrite rules.
It also offers a visual representation of rewriting traces as a derivation tree. The strategy
language is used in particular to guide the construction of this
derivation tree.  The implementation uses the small-step operational
semantics of the language.  Some of these steps require a copy of the
strategic graph program; this is done efficiently in PORGY thanks to the cloning
functionalities of the underlying TULIP system~\cite{auber:2012:hal-00659880:1}.
Verification and debugging tools for avoiding conflicting rules or
non-termination  are planned for future work.

\bibliographystyle{eptcs}
\bibliography{bib}

\end{document}